# Comprehensive recovery of a weak aftershock sequence in the North Atlantic using waveform cross correlation

Dmitry Bobrov, Ivan Kitov, Mikhail Rozhkov, and Gadi Turiomuruguendo

Comprehensive Nuclear-Test-Ban Treaty Organization

Abstract

We apply cross correlation between multichannel seismic waveforms as a technique for signal detection and automatic event building at the International Data Centre (IDC). This technique allows detecting signals with amplitudes by at least a factor of two lower than those found in the current version of IDC processing.Previously, we processed with a cross correlation detector aftershock sequences of a large earthquake with thousands of aftershocks detected by the International Monitoring System (IMS) and a middle-size earthquake (hundreds of aftershocks). Our study has revealed that the official Reviewed Event Bulletin (REB) of the IDC misses from 50% to 70% valid seismic events. Since the IDC is a major contributor to the International Seismological Centre (ISC) these extra events together with the associated arrivals are missing from the ISC bulletin which is an open data source for the broader seismological and geophysical community.Here, we assess the ultimate resolution of the cross correlation technique with specific IDC constraints. The aftershock sequence of the October 5, 2011mb(IDC)4.2 earthquake in the North Atlantic is an example of a weak sequence and includes only 38 REB events.The number and quality of these REB events,which are used as master events, allow conducting a comprehensive interactive reviewby experienced analysts of all event hypotheses obtained by the cross correlation technique.In aniterative procedure starting from the main shock, all 38 REB events were found andanalysts added 26REB events. Therefore, the cross correlation pipeline reduces the detection threshold by a factor of 2 to 3 and approximately doubles the number of events in the REB, and thus, in the ISC bulletin for the North Atlantic.

# Introduction

The completeness of global seismic catalogues, which are a principal part of seismological knowledge, can be improved by development of the global seismological network with more sensitive stations and by applying new processing methods to archive data. We demonstrate that re-processing of multichannel waveforms from array stations of a global network using cross correlation with signals from known events may result in significant improvement of the catalogue completeness.

The International Monitoring System (IMS)of the (Provisional) Technical Secretariat for the Comprehensive Nuclear-Test-Ban Treaty Organization(CTBTO) is a global network designed to provide a uniform detection and monitoring thresholds. Seismic data from 50 primary IMS stations will be continuously transmitted to the International Data Centre (IDC), which has to produce a high quality event bulletin. To enhance the network capability to locate events and to estimate their magnitudes 120 auxiliary stations will be sending data to the IDC on request (Coyne *et al*., 2012). The IMS is not complete yet but many primary and auxiliary stations have been in operation since 2000 creating an extensive archive of original waveforms and bulletins. Using all available stations, the IDC produces a Reviewed Event Bulletin (REB), which in a few months after its official issuance becomes available for the seismological community via the International Seismological Centre. Therefore, the quality of the REB directly affects the completeness of the ISC catalogues, especially, in remote areas not covered by regional networks. The North Atlantic is one of such areas where many ISC events are unique to the IDC as reporting agency.

Cross correlation between seismic waveforms has been introduced and tested as a potential technique for enhancement of signal detection and improvement of automatic event building at the IDC (Bobrov *et al*., 2012a). The physical intuition behind more effective signal detection is simple – small and mid-sized seismic events close in space should produce similar signals as recorded by seismic stations of the International Monitoring System. Equivalently, these signals have to be characterized by higher cross correlation coefficients. When similar seismic signals are mixed with the ambient microseismic noise of the same amplitude the level of correlation between these signalsfalls. Nevertheless, detection of extremely weak signals, even below noise level, is still possible because correlation with the ambient non-coherent noise is asymptotically approaches zero for longer templates. In practice, cross correlation plays a role of matched filter (*e.g*., Van Trees, 1968; Harris, 2006), which is able to find signals below the noise level.

For regional seismic waves, the improvement in detection has been demonstrated for several aftershock sequences within continents. Israelsson (1990) applied cross correlation to high-frequency signals from industrial blasts and showed that carefully selected waveform templates reduce the detection threshold and provide a very sensitive filter for discrimination. Joswig (1990) developed a detection algorithm based on matching of earthquake specific patterns in sonograms. Harris (1991) successfully applied cross correlation to signals from quarry blasts, which are characterized by complex source-time functions. Joswig and Schulte-Theis (1993) applied dynamic waveform matching with adjustable template shape to mining-induced seismicity. These studies proved that the similarity between waveforms generated by spatially close events is high enough to make cross correlation detection by far more efficient than standard energy detectors like the ratio of a short-term average (STA) anda long-term average (LTA).

The next wave of interest to cross correlation of regional seismic phases rose in the beginning of 2000s and was caused by intensive relocation of small and intermediate aftershock sequencesusing the advantage of accurate picking of similar phases (Schaff *et al*., 2004; Schaff and Richards, 2004ab; Schaff and Waldhauser, 2005; Richards *et al*., 2006;

Slinkard *et al*., 2013). With the relocation, a few new problems emerged and old issues were revised. Harris and Paik (2006) and Harris (2006, 2008) proposed to use sub-space detectors instead of original templates in order to enhance detection without significant increase in the number of templates. Gibbons and Ringdal (2004) used array stations NORES, NORSAR, and Hagfors to detect extremely small signals from small, cavity-decoupled, underground explosions in Central Sweden and reported substantial improvement on the beamforming technique. The smallest event was found only when individual cross correlation traces were stacked according to the standard beamforming procedure and STA/LTA detector was applied to the aggregated CC-trace. No signal from the smallest event was visible in the original waveforms. Extending the use of array stations, Gibbons *et al.* (2005) proposed to monitor specific regions with cross correlation technique in order to reduce the detection threshold and the rate of false alarms from events in neighbouring areas. Gibbons and Ringdal (2006) introduced a new measure of magnitude based on cross correlation coefficient. Schaff and Richards (2011) demonstrated that this magnitude measure is characterized by a much lower uncertainty. All these and many other methods and achievementswere directly used for seismic monitoring of nuclear tests (Selby, 2010; GibbonsandRingdal, 2012; Schaff *et al.,* 2012; Bobrov *et al.,* 2012a).

Here, we apply cross correlation to teleseismic waveforms from events in a region not covered by regional networks. At teleseismic distances, seismic wavefield is much simpler than at regionals and local ranges. The P-wave signals are shorter and characterized by a narrower frequency band. The complexity of teleseismic waveforms as defined by the time-bandwidth product (Burnaby, 1953) is much lower than that for regional and local signals. At 3-C stations, the signals from events in different parts of the Earth can also correlate quite well. Hence, the advantage of having higher cross correlation coefficients only for signals from close sources disappears at teleseismic distances (Gibbons and Ringdal, 2012). The use of cross correlation for the P-waves between 20º and 100º would not be effective without arrays stations. Due to the spatial separation of individual sensors of an array an incident planar P-wave arrives at individual channels with varying time delays, which depend on azimuth and slowness of the incident wave and the array configuration. In a way, these time delays are equivalent tothe travel time differences between regular regional phases (*e.g*., Pn and Lg) quickly changing with the separation between sources. Larger aperture provides higher complexity and thus increases the selectivity and reliability of detectionsat a given threshold.

The completeness of the IDC Reviewed Event Bulletin (REB) is a strict requirement of the Comprehensive Nuclear Test Ban Treaty. Currently, the REB is built by automatic and interactive processing. The former creates a standard event list (SEL) based on automatic detections. IDC analysts use all SEL hypotheses and automatic detections to build the REB in line with the requirements of high quality of detections and within the predefined uncertainty limits of arrival time and vector slowness. Analysts can also add detections, which were missed in automatic processing. Accordingly, all detections and events built by the cross correlation technique were reviewed by IDC analysts who have built the final cross correlation bulletin.

The objective of this study is to demonstrate the superior performance of waveform cross correlation in signal detection and event building. When recovering an aftershock sequence, the cross correlation technique provides a complete earthquake catalog to the extent IMS seismic data allow.

## Data and method

For this study, we selected a relatively short aftershock sequence of the October 5, 2011 earthquake in the North Atlantic. Seismic stations of the International Monitoring System detected signals from these aftershocks and, after interactive review of automatic event hypotheses, analysts of the International Data Centre built events (for details of routine processing see, Coyne *et al.*, 2012). The Reviewed Event Bulletin of the IDC contains 34 aftershocks with the main shock coordinates 57.9526ºN 32.5197ºW and the origin time 23:02:10 UTC. The hypocenter of this earthquake was fixed to 0 km and it had magnitude mb(IDC)=4.23. The main shock was not the largest event in the sequence. The earthquake at 00:39:31 on October 6 (57.9122ºN 32.6475ºW) had mb(IDC)=4.79. Also, there were three smaller foreshocks approximately one hour before the main shock which we also included in our analysis together with their time block. All aftershocks were detected within ~26 hours and then the sequence ceased. To confirm this observation, we processed extra ten hours using cross correlation detector. Therefore, the total length of the analyzed interval was 36 hours. The relevant continuous data and waveform templates were retrieved from the IDC archive database. The size and duration of the sequence allowed interactive reviewing all new event hypotheses generated by cross correlation.

For cross correlation analysis we use only those array stations of the Primary Seismic Network of the IMS which detected the P-waves. We exclude primary 3-C stations from the analysis because they do not provide the appropriate accuracy of azimuth and slowness estimates for the P-wave detections obtained by cross correlation. Several auxiliary IMS array stations, which detected the P-waves from the studied sequence, were also excluded since they do not provide continuous waveforms for the studied period. Auxiliary stations send short data segments by request initiated by event hypotheses built using the Primary Seismic Network.

In total, 21 array stations reported at least one P-wave arrival from this sequence. Figure 1 displays the positions of these stations relative to the main shock. After a careful inspection of station sensitivity to earthquakes in the North Atlantic and the quality of the REB detections, only 10best stations were selected for creation of waveform templates and further continuous cross correlation. Other 11 stations are only able to detect the largest events and thus worthless for detection of the weakest signals. The selected 10 stations are highlighted in Figure 1. Four of them (ILAR, PDAR, TXAR, and YKA) are in North America, five (AKASG, BRTR, GERES, MKAR, and SONM) in Eurasia, and one (TORD) is in Africa. The distribution of these ten stations is characterized byan azimuthal gap of ~180°.

We used automatic cross correlation processing pipeline, which had been developed at the IDC for monitoring purposes (Bobrov *et al.*, 2012a), and built a cross correlation standard event list, XSEL. For correlation, several template waveforms recorded by primary IMS array stations were preselected and then correlated with the continuous data at the same stations. A valid detection has two principal features: the absolute value of cross correlation coefficient and the signal-to-noise ratio (SNR) estimated from the aggregate CC-traces should be above station specific thresholds. These detections are associated together to create XSEL event hypotheses using the estimated origin times, i.e. the arrival times less the relevant master event/station travel times.

The quality of REB events is guaranteed by comprehensive human testing – each and every event and detection is checked by at least two analysts. To ensure the same quality for the detections and events obtained by cross correlation an experienced analyst has reviewed all XSEL hypotheses and associated detections according to standard IDC rules and guidelines. To create REB capable events, the analyst was alsoallowed to add appropriate detections at primary 3-C stations and auxiliary seismic stations and to determine standard characteristics of all signals: arrival time, amplitude, period, and vector slowness together with their uncertainties.

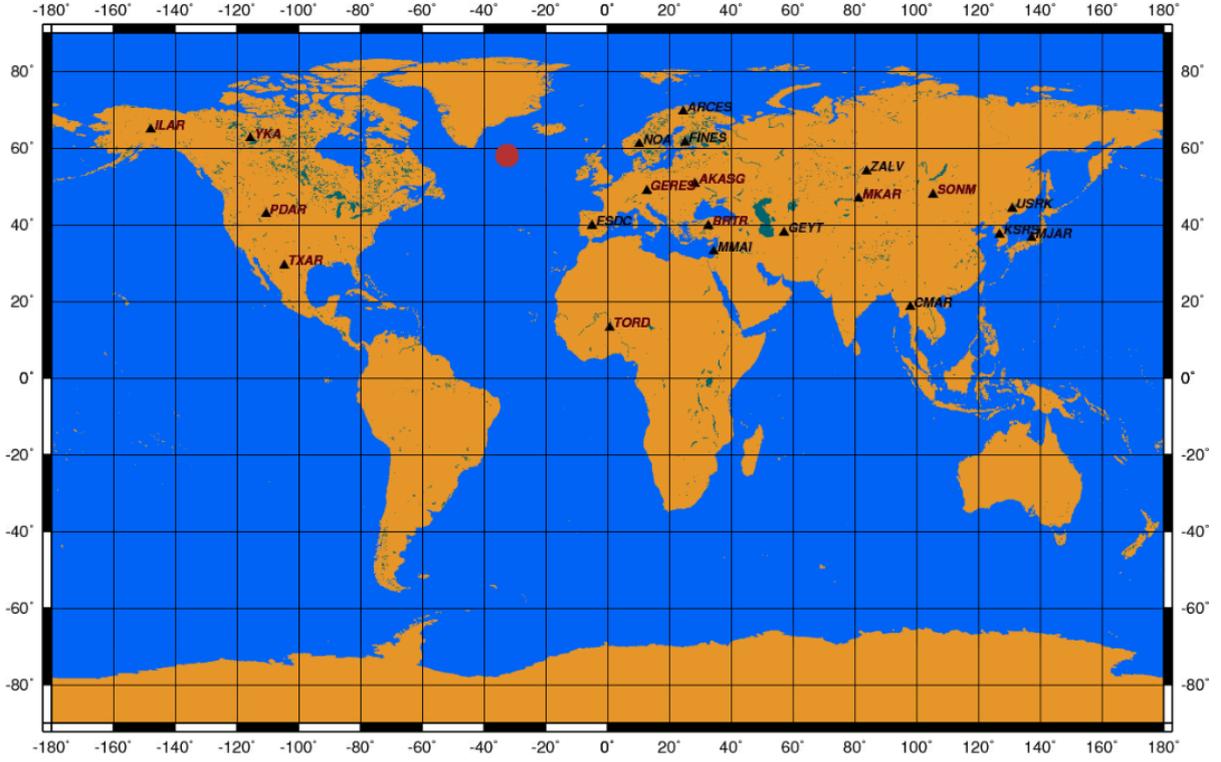

Figure 1. The main shock (red circle) and distribution of IMS arrays. Ten selected arrays are highlighted red.

The processing pipeline starts with calculation of cross correlation coefficient. For array stations, Gibbons and Ringdal (2004, 2006) proposed the use of a normalized cross correlation function averaged over all channels. It is assumed that master events and continuous time series have the same sampling rate at all channels. With permanent progress in seismic instruments as well as in methods of data acquisition and processing, seismic stations are upgraded every five to ten years. Cross correlation requires that the absolute positions of instrument sites in a given array and the amplitude-frequency characteristics of measuring channels should not change over time.

For an individual analogue channel $j$ measuring a continuous time function $x_j(t)$, we define a discrete template time series consisting of $N$ consecutive samples using the notation $\mathbf{x}_{jn}(t_0)$, where $t_0$ is the absolute time of the template signal. The samples should be taken at a constant sampling rate, $\Delta t$. For the same channel, the continuous waveform $y_j(t)$ has the same sampling rate and the normalized cross correlation coefficient between $\mathbf{x}_{jn}(t_0)$ and $\mathbf{y}_{jn}(t)$ at absolute time $t$ is defined by the following relationship:

$$CC_j(t) = \mathbf{x}_{jn}(t_0) \cdot \mathbf{y}_{jn}(t) / (||\mathbf{x}_{jn}(t_0)|| \cdot ||\mathbf{y}_{jn}(t)||) \qquad (1)$$

where $||\cdot||$ denotes the L2 norm of the relevant time series of $N$ samples. The cross correlation coefficient for the given channel is a discrete time series with the same sampling rate as the original signals. To calculate an aggregate cross correlation coefficient we average the individual normalized cross correlation functions over all channels:

$$CC(t) = \sum_{j=1}^{M} CC_j(t) / M \qquad (2)$$

Where *M* is the number of channels in the array. The averaging is flexible to exclusion of damaged channels and allows avoiding many problems with data quality such as spikes, gaps, high noise at individual channels (Bobrov *et al.,* 2012a). As a result, the averaged cross correlation trace has no artificial steps associated with a sudden change in the number of working channels, which might be misinterpreted as a valid detection. More important, averaging of coherent CC traces improves the signal-to-noise ratio more effectively than standard beam forming. Gibbons and Ringdal (2004) were likely the first to report the superiority of the aggregate CC over beam forming as a detection tool.

There is one technical problem with (1), which may reduce the true value of the aggregate CC. The same P-wave arrives at individual channels of a given array at different times. For a waveform template, relationship (1) implies that varying parts of the sample signal will be used at different channels since all template signals start at the same absolute time $t_0$. We settle this problem by shifting all channels by their theoretical delay times relative to the reference station, where the absolute arrival time is measured. For a given master/array station configuration one has the following relationship for individual time delays:

$$dt_j = S_n \cdot dnorth_j + S_e \cdot deast_j \tag{3}$$

where $dt_j$ is the theoretical time delay of the arrival at the *j*-th channel relative to the reference channel of the array, $S_e$ and $S_n$ are the east-west and north-south components of the vector slowness defined by the master/station pair, $deast_j$ and $dnorth_j$ are the east-west and north-south coordinates of the *j*-th array element relative to the reference station. Relationship (3) implies perfect planar propagation. Using (3) one can rewrite (1)

$$CC_j(t) = \mathbf{x}_{jn}(t_0 \text{-} dt_j) \cdot \mathbf{y}_{jn}(t\text{-}dt_j) / (||\mathbf{x}_{jn}(t_0\text{-}dt_j)|| \cdot ||\mathbf{y}_{jn}(t\text{-}dt_j)||) \tag{4}$$

In (4), all template signals at individual channels of the array span practically the same segment of the true signal. The absolute arrival times at individual channels still may differ by small empirical time delays associated with non-planar propagation on the incident P-wave due to the inhomogeneous velocity structure beneath the array. The continuous waveforms, $\mathbf{y}_{jn}(t\text{-}dt_j)$, are also shifted by the same theoretical time delays and thus retain the empirical time delays for slave events in the near proximity of the master. This is an important advantage of cross correlation over standard beam forming, where the empirical time delays are not compensated, desynchronize signals in stacked channels, and result in beam loss (Coyne *et al.,* 2012).

To improve the existent bulletins, we are searching low amplitude signals from various types of sources (*e.g.,* underground explosions and earthquakes) at teleseismic distances. In many cases, these signals are so difficult to find in raw waveform data that standard detection methods miss them. The weakest signals of interest are always short and barely above the ambient noise level. Therefore, all templates in our study include only a few seconds of P-wave signals and a short time interval before the signal (lead).

Frequency filtering is a well-known tool to improve SNR, and thus, to enhance detection. All waveforms from individual vertical channels of IMS array stations are filtered using a set of four causal band-pass filters. The frequency band defines the length of template window. For example, the template for the low-frequency (Butterworth, order 3) filter between 0.8 Hz and 2.0 Hz is 6.5 s long including 1 s before the arrival time. Table 1 lists characteristics of four templates used in this study.

When an aggregated CC time series is calculated for a given time interval one can apply signal detection algorithms. By design, cross correlation coefficient can be used for detection as it is by introduction of a station/master specific threshold, $CC_{tr}$, which should be determined empirically. These thresholds are likely to vary with characteristics of ambient noise, seismic phase (*e.g.*,Lg, Pn, and P), station configuration (3-C or array), station/master propagation path including local velocity structures in the source region and under the station, source function (*e.g.,* underground explosion for seismic monitoring under the CTBT), and source depth.

Table 1.Time windows and frequency bands of the templates for P-wave arrivals.

| Filter | | | | Window, s | | |
| --- | --- | --- | --- | --- | --- | --- |
| Low (Hz) | High (Hz) | Type | Order | Lead | Signal | Name |
| 0.8 | 2.0 | BP | 3 | 1.0 | 5.5 | P0820 |
| 1.5 | 3.0 | BP | 3 | 1.0 | 4.5 | P1530 |
| 2.0 | 4.0 | BP | 3 | 1.0 | 3.5 | P2040 |
| 3.0 | 6.0 | BP | 3 | 1.0 | 3.5 | P3060 |

For weak signals, the absolute value of correlation coefficient for even identical events can be reduced by the effect of uncorrelated seismic noise mixed with these signals. Instead of using *CC* as a detector we make use of the energy detector, which is already implemented at the IDC for original waveforms. This detector is based on a running ratio of a short-term-average (STA) and a long-term-average (LTA), which is computed recursively using previously computed STA values. The LTA lags behind the STA by a half of the STA window. After a thorough investigation we determined the following windows: 0.8 s for the STA and 40 s for the LTA. These windows take into account the narrow frequency band and short time windows used in this study for the P-waves: the aggregate CC traces may contain spikes associated with noise due to the limited time–bandwidth product (Gibbons and Ringdal, 2006; Schaff and Richards, 2011). To iron these spikes out the STA is approximately equal to one period associated with the lowermost frequency among all used filters - 0.8 s. In our study, a valid signal is detected when STA/LTA ($SNR_{CC}$) is above 3.0. It is worth noting that for original waveforms a valid signal usually has SNR>2.0, but the CC detector can find valid signals with standard SNR of 1.0 and even lower (Schaff and Richards, 2011; Bobrov *et al.*, 2012b).

We apply the STA/LTA detector to aggregated CC traces at all stations associated with a given master-event. A set of qualified CC-detections is obtained with their arrival times, $t_{km}$, where *k* is the index of the *k*-th arrival at station *m*. For further analysis, we assume that all valid arrivals detected by cross correlation should belong to some events near the master. These are the slave events we are seeking for. Due to the master/slave spatial closeness, the travel times from these sought events to the relevant stations can be accurately approximated by the master/station travel times, $tt_m$. Then the origin times associated with the obtained detections are the difference between their arrival times and the approximated travel times (same for all events around the master one)

$$ot_{km} = t_{km} - tt_m \qquad (5)$$

Now we have a set of origin times likely associated with slave events located within tens of kilometers around the master event. Origin time is a natural characteristic of any seismic source. In the simplest approach, each and every origin time could be considered as defining a unique event. When a few origin times at different stations are very close in absolute time (say, within 6 sec) they are characterized by a non-zero probability to be associated with physically the same slave event. Depending on the number and quality of associated arrivals the obtained slave events have a varying reliability. In order to avoid the uncertainty in the definition of a reliable event the IDC has adopted a set of quantified quality criteria called "event definition criteria" (EDC). One needs three or more primary IMS stations to "build" a seismic event. In addition, the detections associated with this event have to match strict measurement uncertainty bounds (Coyne *et al*., 2012).

For the whole multitude of origin times obtained by cross correaltion at ten selected stations, the task is to unambiguously distribute them among slave events. This process is called "local association", *LA*, in line with the name of global association, *GA*, which is a set of applications currently used by the IDC for automatic event building (Coyne *et al*., 2012). Indeed, only the phases from slave events local to the master one should be associated. The *LA* does not "see" any events around the master event beyond the radius of correlation. The accuracy of master/station travel time prediction for the slaves detected by cross correlation degrades with the master/slave distance. For the studied aftershock sequence, the master/slave distance may reach 50 km with the travel time differenceof a few seconds. For stations on the opposite sides of the master the total travel time difference, and thus, the origin time bias can be 5 and more seconds.

In order to remove this bias and to improve the process of origin time association with a unique slave event we have introduced an equidistant grid around each master event. There were 6 and 12 points evenly distributed over two circles of 25 km and 50 km in radius, respectively. In total, one master generates 19 nodes. For each node, the same arrival times obtained by cross correlation are reduced to origin times using the theoretical travel times corresponding to the node location. When the arrival times are accurately estimated, the smallest RMS origin time error is likely obtained for the node closest to the true location of the sought slave event. The set of origin times with the lowermost RMS error defines a tentative slave event hypothesis. Its origin time is calculated as the median origin time for all associated arrivals and the best node is considered as the slave location. The solutions for all other nodes are erased.

The detections obtained from the aggregate CC traces are characterized bythe estimated arrival time, *CC*, and $SNR_{CC}$, which can be used to assess the overall quality of signals. These signals can be also characterized by various dynamic and kinematic parameters useful for phase association and event creation. In this study, we use the estimates of azimuth and slowness from the multichannel CC traces as well as the relative slave size.

Gibbons and Ringdal (2006) proposed and successfully applied *f-k* analysis to the cross correlation time series at array stations. This allowed a significant improvement in the overall resolution due to the sensitivity of correlation to the distance between events and effective suppression of incoherent noise. We have estimated pseudo-azimuth and pseudo-slowness using *f-k* for the correlation time series. The term "pseudo" is used since there is no one-to-one correspondence between the ground motion and CC domains, and the estimated azimuths and slownesses are not expressed in genuine degrees and seconds per degree. Nevertheless, the azimuth and slowness residuals obtained from CC traces depend of the master/slave distance and effectively reject most of cross correlation detections associated with noise and remote events (Gibbons and Ringdal, 2012). We apply the following thresholds to (pseudo) azimuth and slowness residuals: 20° and 2 s/deg, respectively. All detections with larger estimated residuals are removed from the detection list used for the *LA*.

Bobrov *et al.* (2012a) proposed a dynamic variable for rejection of invalid arrivals. It is based on the ratio of the L2 norms of two signals:$||\mathbf{s}||/||\mathbf{m}||$, where **s** and **m** are the vector data of the slave and master, respectively. The logarithm of the ratio

$RM = \log(||\mathbf{x}||/||\mathbf{y}||) = \log||\mathbf{x}|| - \log||\mathbf{y}||$ (6)

is the size difference between two signals or their relative magnitude. This difference has a clear physical meaning for close events with similar waveforms. For a given slave/master pair characterized by similar source functions and propagation paths, the relative magnitude should not vary much over detecting stations. Larger deviations from the relative magnitude averaged over all involved stations can be interpreted as an indication of wrong or irrelevant detection. This observation provides a reliable dynamic parameter for a correct arrival association at several stations. In this study, we have adopted the threshold of 0.7 for the station *RM* deviation from the network average.

The *LA* is a simplistic process compared to the Global Association. The principal advantage of the *LA* is a significantly reduced number of arrivals for a given master event from the associated stations. At the same time, the total number of arrivals at a given station may double relative to that from the current IDC detector.

When several master events are close in space, their templates may correlate with waveforms from the same event, and thus, create similar new event hypotheses with very close arrival times at several stations associated with these masters. To select the best one from a few similar events with close arrival and origin times we count the number of stations used in these hypotheses. If one event has the largest number of stations it is retained as an XSEL event. When several events have the same largest number of stations we select the one with the smallest standard deviation in origin times. By definition, this event is the most reliable and its parameters are written into the database. All competing hypotheses are rejected and the associated detections removed from the relevant detection lists. Thus, for a multiple set of master events, the *LA* provides a unique set of event hypotheses.

## Results

The main shock on October 5, 2011 was detected by 21 array and 9 3-C stations. From 21 IMS arrays only ten most sensitive to this specific region were used in our study. The other 11 arrays were able to detect only the main shock and the biggest events in the aftershock sequence. In the process of event creation, their value added to is close to zero and thus they can be excluded without loss in resolution and sensitivity.

Thirty six hours of continuous data were processed using waveform templates from the selected stations with increasing number of master events. For the first iteration, waveform templates were selected from the main shock only. After the first iteration, 54 REB capable events were created by analyst from 84 XSEL hypotheses. This set included 34 actual REB events, as was confirmed by arrival times of defining phases at the selected IMS stations.

From 54, only 40 events have signals at three or more primary array stations with high enough quality to be used as masters. These 40 events were used in the second iteration as master events, and the same 36-hour IMS data were processed again. Only six new, i.e. not found in the first iteration, were found, with only 3 of them REB compatible. They all were small and could not be used as masters. The iteration stopped. The final XSEL included 64 events, i.e. 67% more than the REB for the same sequence. It that sense, the XSEL is a more complete catalog.

The overall performance of waveform cross correlation depends on the difference between the detections associated with the XSEL and those which are not associated. We have studied some of defining characteristics in detail. They are as follows: relative performance of detection beams; relative performance of stations; arrival time residuals between the REB and XSEL; frequency distribution of cross correlation coefficient as a measure of probability for a detection to be associated; frequency distribution of relative magnitude; frequency distribution of cross correlation SNR; all these characteristics for individual stations

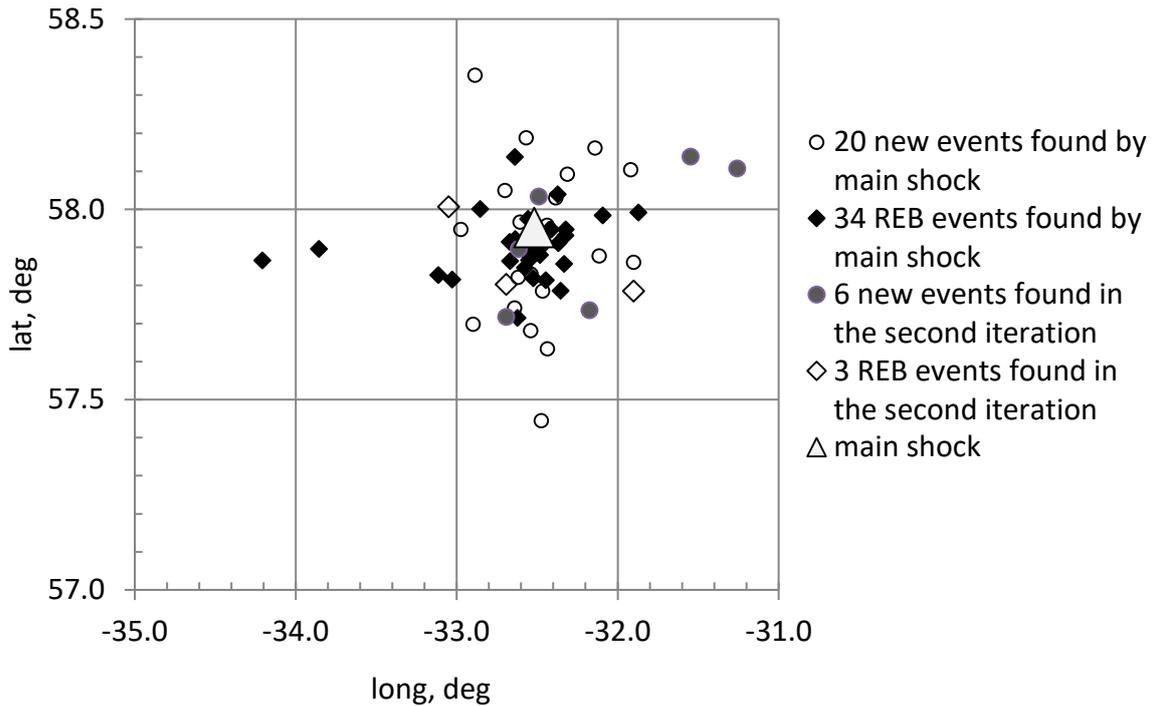

Figure 2. Locations of aftershocks and the main shock

The above map illustrates the iterative process of the complete recovery of the sequence. The main shock was used as a master event. This is the first event in the sequence and big enough to have clear signals at many primary stations. (We recommend the use of main shocks as master events in all cases when their magnitudes are less than 5.5.) The first master has found 34 from 37 (38 less the main shock) REB events and also 20 new events, i.e. 54 events in total. There were 24 events in the XSEL and the analyst considered the remaining 4 events as marginal because their signals were not visible. (Notice, waveform cross correlation can be very sensitive and detect signals below the noise level.) In order to select high quality master events for this seismic region we studied separately the statistics of detected and associated (with events) P-waves. We also evaluated the performance of stations and detection beams to tune the detection and phase association process.

The events in the studied aftershock sequence were detected by cross correlation using various beams. The nature of their source mechanism implied slow rupture and the relative enhancement of low frequency content. As expected, the P0820 CC-beam gave the largest number of detections (6144) used in all events built with cross correlation. Surprisingly, the highest frequency CC-beam, P3060, gave 40 detections. For ARCES it is a natural frequency band for detection from small events. For other stations, high frequency detections might be related to higher noise level at lower frequencies may be wrong. In any case, we would not recommend removing the high frequency CC-beam from cross correlation processing. Other two CC-beams detected 3426 (P1530) and 1284 (P2040) signals.

It is a well-known fact that IMS seismic stations are characterized by large differences in the total number of detection and the number of those used in the REB. The input of a station depends on its sensitivity and resolution. The waveform cross correlation technique reveals the same general tendency. Figure below shows the portions of all detections and those used in the XSEL for all stations. The largest portion of valid detections is provided by TORD. It has a high overall resolution and is also situated at a fortunate distance. ILAR gives the largest portion of false detections but is also indispensable for the XSEL. MKAR and AKASG are prolific with low false alarm rate. There are quite a few stations which should not be used in the study at all because their value added is negligible. For example, PETK gives only one valid detection for the biggest event. The best ten stations are: AKASG, BRTR, GERES, ILAR, MKAR, PDAR, SONM, TORD, TXAR, and YKA.

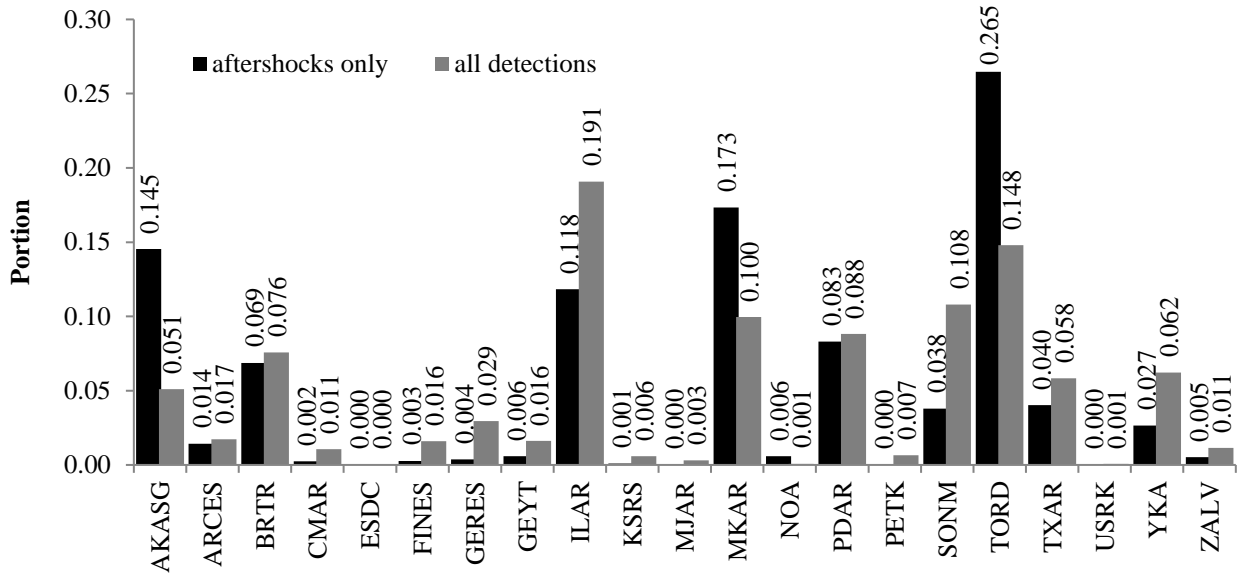

Figure 3. The distribution of the portion of detections at 21 IMS array stations as estimated from all cross correlation detections and those associated with XSEL events.

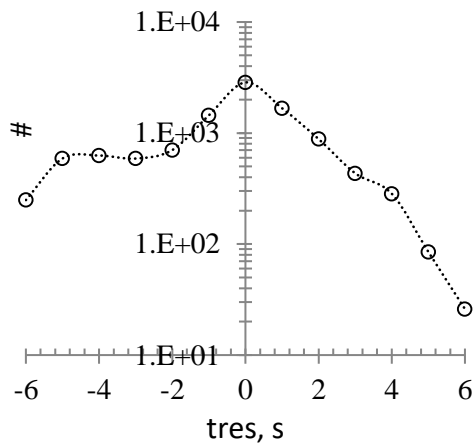

Figure 4. Frequency distribution of the travel time residuals, $t_{res}$, for the associated cross correlation detections. Notice the lin-log scale.

Frequency distribution of the travel time residuals, $t_{res}$, as determined by the deference between the REB arrival time and that determined by cross correlation for a given master/slave pair. We limit the absolute value of the largest possible residual to 6 s. Notice that the early CC arrivals (positive difference) are characterized by exponential distribution and the late arrivals have a shelf between -2 s and -5 s. Most of arrivals are within ±2s.

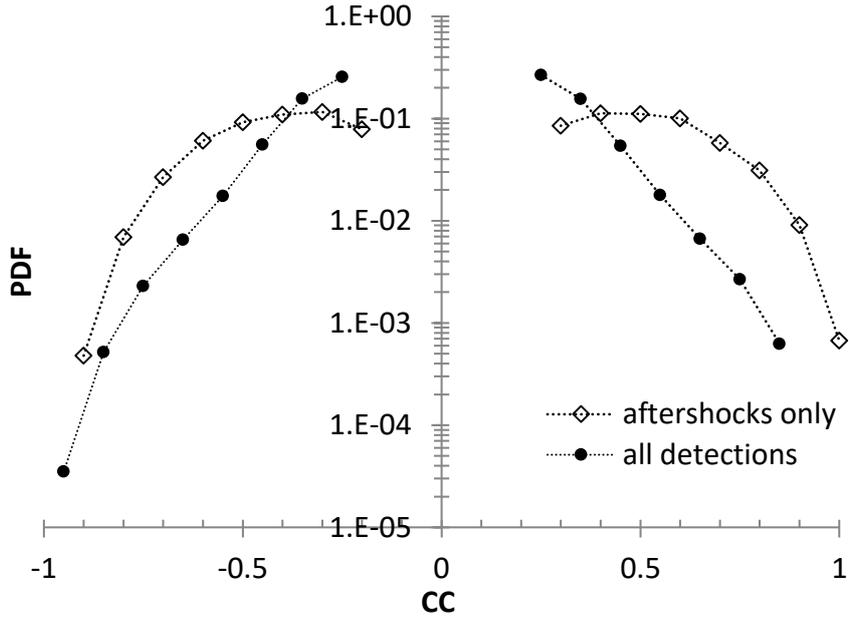

Figure 5. Probability density functions for CC as determined for aftershocks only and all detections. For the XSEL events, both positive and negative CC estimates are normally distributed for |CC|>0.5. Smaller CCs are relatively underrepresented. For all detections, both sides are rather following exponential distributions.

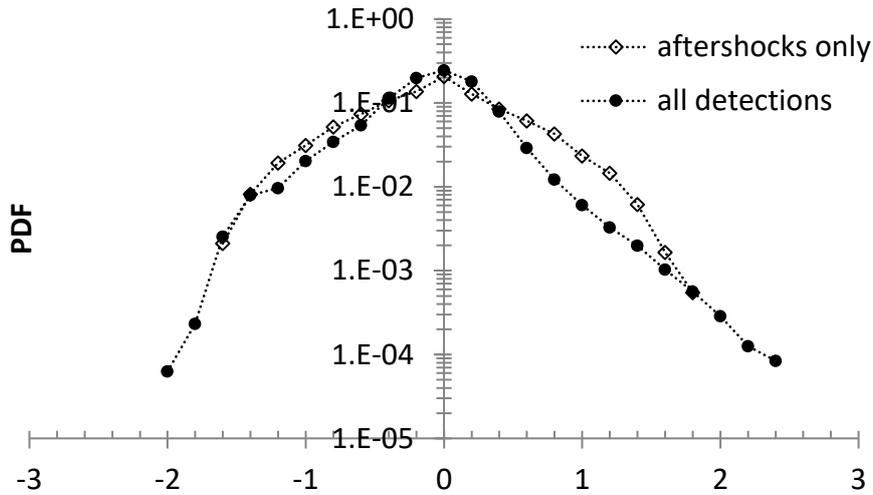

Figure 6. Frequency distribution of *RM* for all aftershocks and all detections. Only 50 aftershocks were used as master events. Notice the lin-log scale. The former curve follows a normal distribution and the latter is closer to an exponential one. The relative magnitudes of the aftershocks are from -1.6 to +1.6. Therefore, the range of relative magnitudes is ~1.6. With the largest event with $m_b$(IDC)=4.51, the smallest has to be ~2.9.

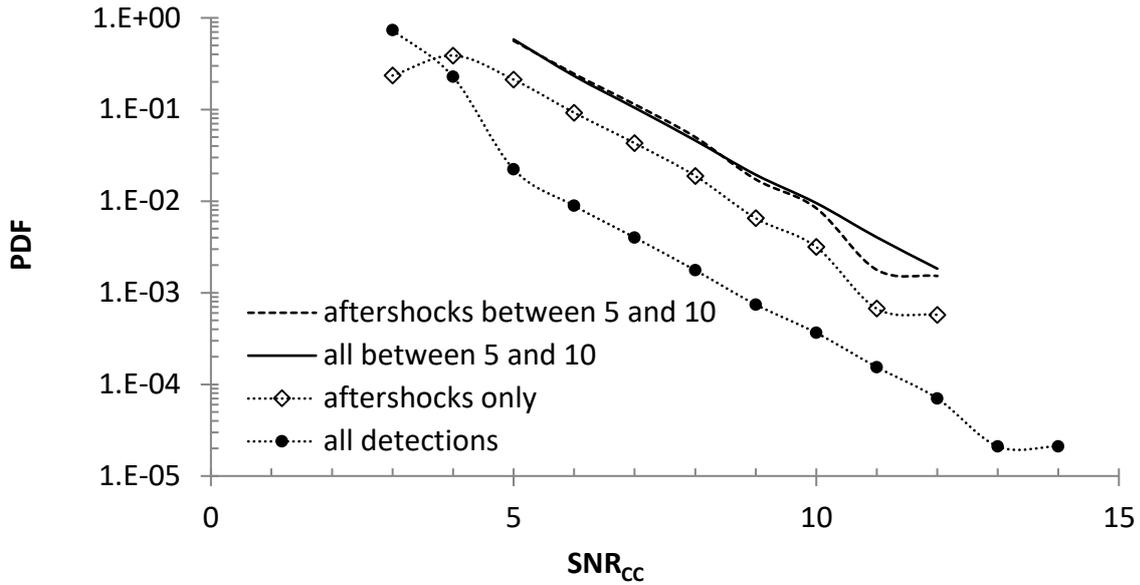

Figure 7. Probability density function for the SNR estimates from cross correlation traces, which were used for detection (SNR>3). Two curves are displayed: for all detections (black circles) and those associated with the XSEL aftershocks (red circles). Both curves clearly follow an exponential law. The curve for aftershocks has exponent of -0.82 for $SNR_{CC}$ between 4 and 10. The curve for all detection follows the same exponential distribution for $SNR_{CC}>5$. When normalized, both curves coincide between $SNR_{CC}$ 5 and 10. Hence, we are likely missing some detections with good SNR which cannot be associated with valid XSEL events. The event definition criteria prohibit events with only two or one detection.

Twenty six new aftershocks were built with cross correlation. They span a 25-hour time interval. Body wave magnitudes range from 3.31 to 4.02. The number of defining phases (including those from auxiliary arrays and all 3-C stations) varies from 4 to 10. The distance from the main shock was calculated using a standard IDC location program. It varies from 4.8 km ($m_b$(IDC)=3.59 and 10 defining arrivals) to 76.7 km ($m_b$(IDC)=3.46 and only 3 defining arrivals). The latter estimate is likely biased characterized by high uncertainty (confidence ellipse).

Table 2. New aftershocks matching the EDC

| Date | Time | $m_b$(IDC) | ndef | Dist, km |
| --- | --- | --- | --- | --- |
| 05/10/2011 | 23:06:39 | 3.32 | 4 | 49.52 |
| 05/10/2011 | 23:51:58 | 3.31 | 5 | 28.16 |
| 05/10/2011 | 23:54:20 | 3.80 | 6 | 38.17 |
| 05/10/2011 | 23:55:51 | 3.65 | 8 | 13.8 |
| 06/10/2011 | 00:01:38 | 4.02 | 10 | 5.36 |
| 06/10/2011 | 00:02:31 | 3.59 | 10 | 4.79 |
| 06/10/2011 | 00:05:23 | 3.56 | 7 | 35.97 |
| 06/10/2011 | 00:10:45 | 3.46 | 3 | 76.62 |
| 06/10/2011 | 00:39:18 | 3.42 | 4 | 36.28 |
| 06/10/2011 | 00:45:54 | 3.52 | 8 | 9.17 |
| 06/10/2011 | 01:07:19 | 3.46 | 8 | 39.36 |
| 06/10/2011 | 02:31:30 | 3.37 | 5 | 31.8 |
| 06/10/2011 | 02:34:16 | 3.35 | 6 | 61.11 |
| 06/10/2011 | 07:36:33 | 3.32 | 5 | 26.43 |
| 06/10/2011 | 07:47:04 | 3.59 | 6 | 56.75 |
| 06/10/2011 | 07:48:25 | 3.54 | 9 | 24.72 |
| 06/10/2011 | 10:37:51 | 3.61 | 5 | 27.02 |
| 06/10/2011 | 13:17:19 | 3.54 | 4 | 11.75 |
| 06/10/2011 | 14:12:58 | 3.38 | 6 | 8.48 |
| 06/10/2011 | 14:17:29 | 3.51 | 8 | 15.83 |
| 06/10/2011 | 19:43:05 | 3.40 | 5 | 30.34 |
| 06/10/2011 | 19:47:26 | 3.61 | 5 | 15.18 |
| 06/10/2011 | 19:57:48 | 3.56 | 6 | 19.81 |
| 06/10/2011 | 21:12:34 | 3.51 | 8 | 32.3 |
| 06/10/2011 | 21:15:49 | 3.46 | 5 | 18.9 |
| 06/10/2011 | 23:49:25 | 3.62 | 9 | 25.47 |

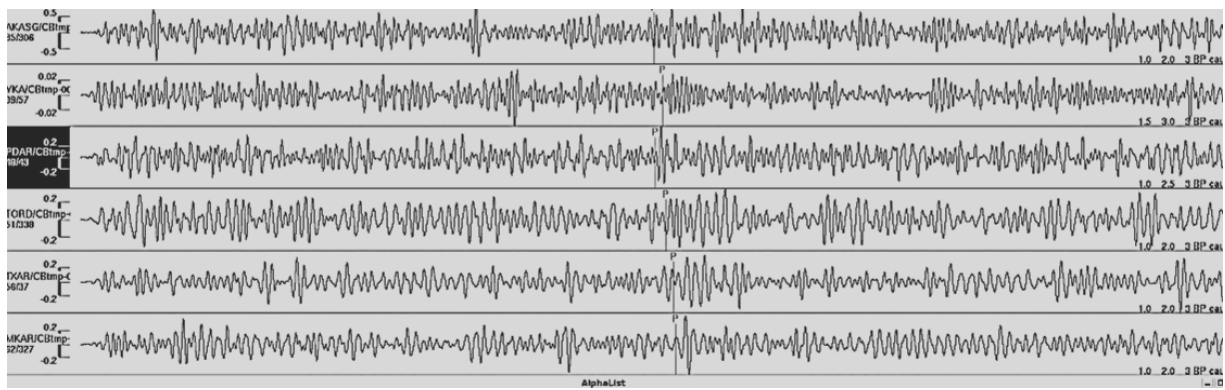

Figure 8. A new (XSEL only) event matching the event definition criteria with mb=3.4. All six defining P-wave arrivals were found by cross correlation at primary array stations. All azimuth and slowness residuals match the predefined uncertainty limits for these arrays.

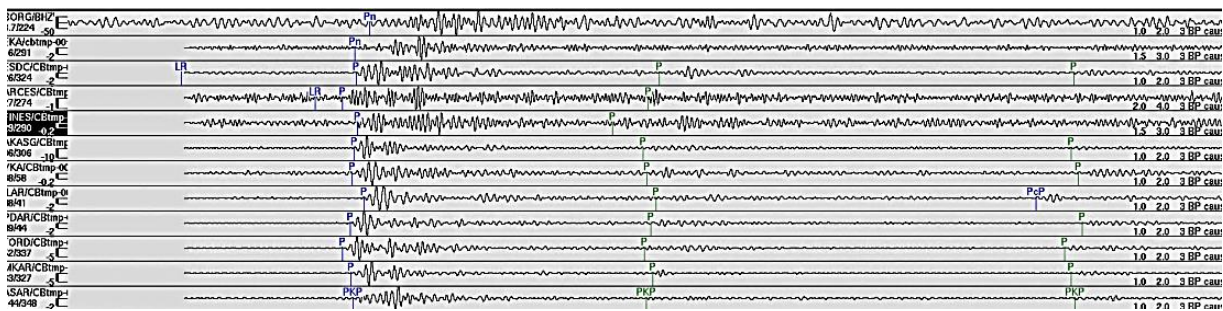

Figure 9. A new (XSEL only) event matching the event definition criteria with mb=4.02. Nine P-wave arrivals were found by cross correlation at primary array stations. Eight of them are time defining and one (FINES) is associated. ASAR has detected PKP. One P-arrival belongs to a 3-C station GNI.

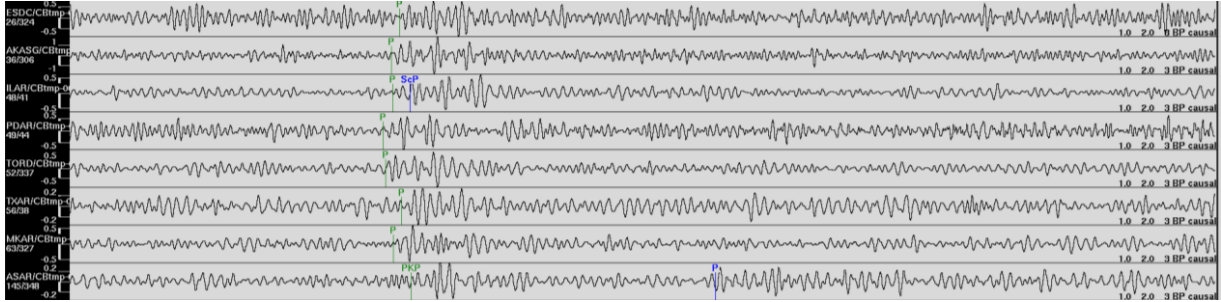

Figure 10. A new (XSEL only – green phases) event matching the event definition criteria with mb=3.5. Seven P-wave arrivals were found by cross correlation at primary array stations.

Table 3. The number of events found by 10 best masters.

| Orid | mb | Total found | REB | New |
|---|---|---|---|---|
| 7980438 | 4.226 | 54 | 34 | 20 |
| 7983356 | 4.127 | 48 | 34 | 14 |
| 7983740 | 4.264 | 44 | 32 | 12 |
| 7990093 | 3.712 | 36 | 28 | 8 |
| 7990652 | 4.186 | 47 | 32 | 15 |
| 7990752 | 3.728 | 38 | 28 | 10 |
| 7982858 | 4.283 | 50 | 35 | 15 |
| 7980722 | 4.755 | 47 | 34 | 13 |
| 7980735 | 4.788 | 47 | 33 | 14 |
| 7990812 | 4.272 | 44 | 34 | 10 |

The main shock was the most efficient master event. There were other events which could serve as masters. Even the event with mb=3.73 has found 28 REB events and 10 new events, as the above Table shows.

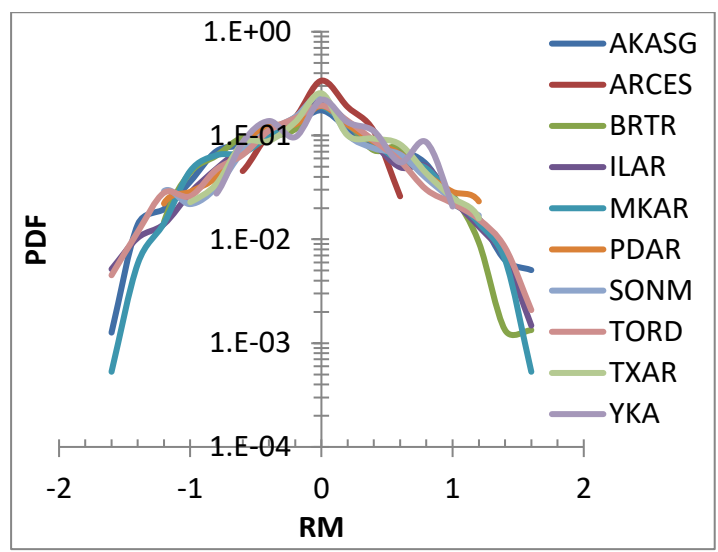

Figure 10. Probability density function of **RM** for ten best primary IMS stations as measured from detections associated with aftershocks. All stations demonstrate similar distributions. Notice the lin-log scale.

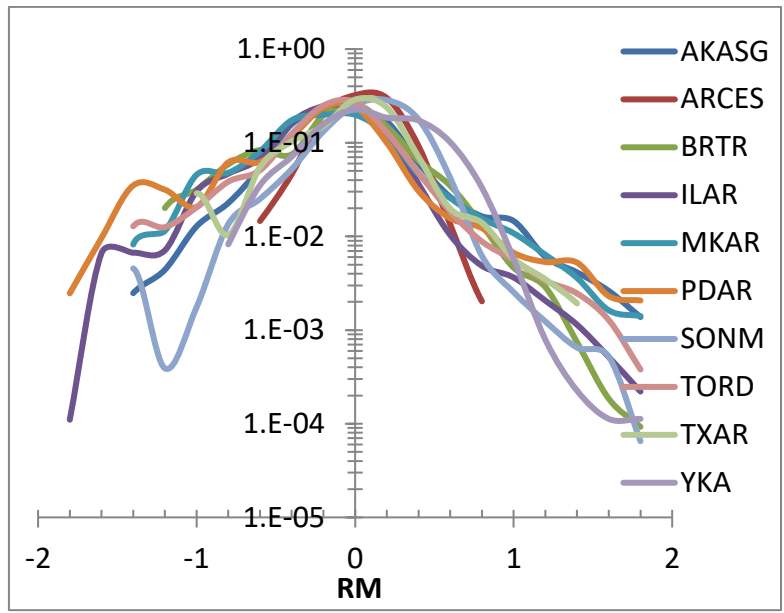

Figure 11. Probability density function of **RM** for ten best primary IMS stations as measured from all detections. Distributions vary from station to station, i.e. some stations produce more bogus detections and wrong **RM** estimates than others.

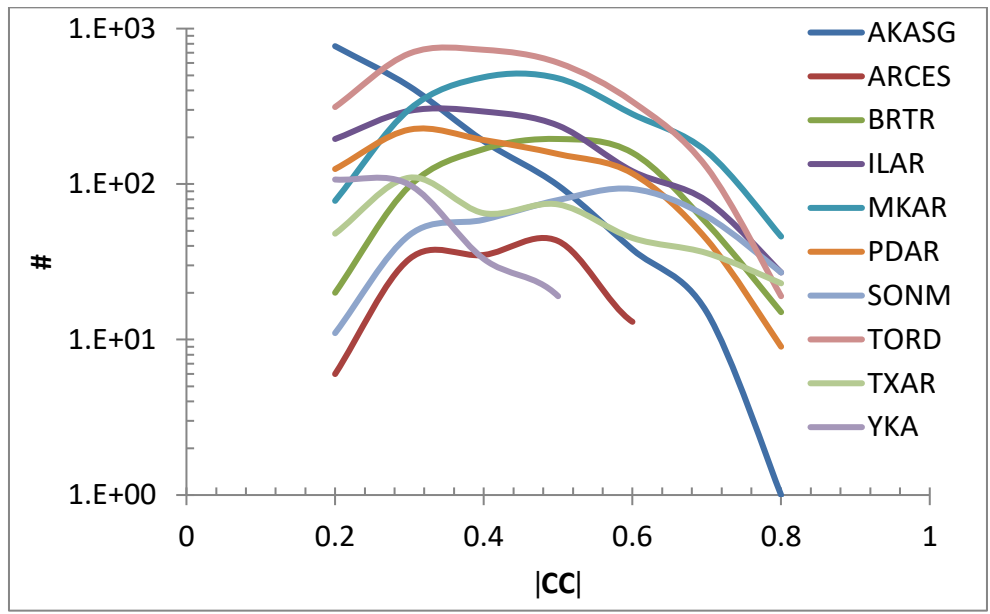

Figure 12. Frequency distribution of CC at ten best primary IMS stations as measured from detections associated with aftershocks. Stations demonstrate quite different performance. These curves may be used to tune station dependent CC-thresholds. Notice the lin-log scale.

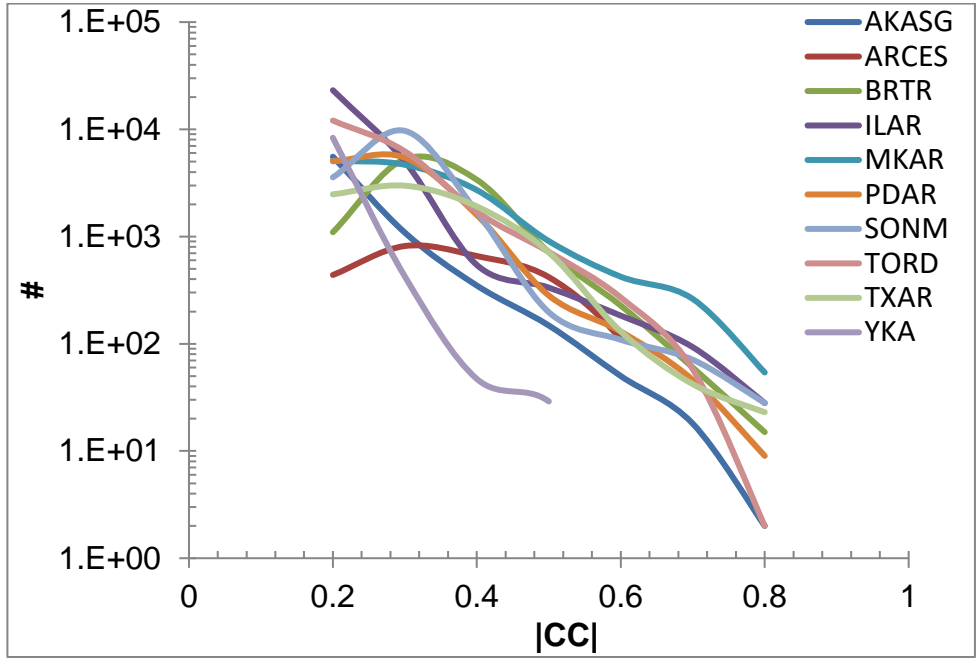

Figure 13. Frequency distribution of |**CC**| at ten best primary IMS stations as measured from all detections. Stations demonstrate similar quasi-exponential distributions.

## Discussion

Any cross correlation coefficient above a predefined threshold can be considered as a signature of a valid signal. With a dense grid of master events with high quality template waveforms at primary array stations of the IMS one can detect signals from and then build natural and manmade seismic events close to the master ones. The use of cross correlation allows detecting smaller signals (sometimes below noise level) than provided by the current IDC detecting techniques. As a result it is possible to automatically build from 50% to 100% more valid seismic events than included in the Reviewed Event Bulletin (REB) of the IDC. We have developed a tentative pipeline for automatic processing at the IDC. It includes three major stages. Firstly, we calculate cross correlation coefficient for a given master and continuous waveforms at the same stations and carry out signal detection as based on the statistical behavior of signal-to-noise ratio of the cross correlation coefficient. Secondly, a thorough screening is performed for all obtained signals using *f-k* analysis and *F*-statistics as applied to the cross-correlation traces at individual channels of all included array stations. Thirdly, local (i.e. confined to the correlation distance around the master event) association of origin times of all qualified signals is fulfilled. These origin times are calculated from the arrival times of these signals, which are reduced to the origin times by the travel times from the master event. An aftershock sequence of a mid-size earthquake is an ideal case to test cross correlation techniques for automatic event building. All events should be close to the main shock and occur within several days. Here we analyze the aftershock sequence of an earthquake in the North Atlantic Ocean with mb(IDC)=4.79. The REB includes 38 events at distances less than 150 km from the mains hock. Our ultimate goal is to exercise the complete iterative procedure to find all possible aftershocks. We start with the main shock and recover ten aftershocks with the largest number of stations to produce an initial set of master events with the highest quality templates. Then we find all aftershocks in the REB and many additional events, which were not originally found by the IDC. Using all events found after the first iteration as master events we find new events, which are also used in the next iteration. The iterative process stops when no new events can be found. In that sense the final set of aftershocks obtained with cross correlation is a comprehensive one.

The iterative procedure based on waveform cross correlation has shown an excellent performance. The aftershock sequence in North Atlantic has been completely recovered using the cross correlation technique with data from the IMS seismic network.

The Reviewed Event Bulletin has been extended by 67%: 26 new events were added to 38 REB events including the main shock. Some of new events have body wave magnitudes above 4.0 and eight primary IMS stations with time defining arrivals. All REB events were also found.

The main shock is an excellent master event, which has found 54 from 64 events (84%) in the final XSEL. Several larger aftershocks from the sequence could also be used as master events with high performance .

The frequency distribution of cross correlation coefficients,**CC**, for signals used in all built events has two broad peaks around +0.5 and -0.5.  A larger **CC** provides higher probability for a signal to be associated with a valid XSEL event.

The relative magnitude**, RM**, introduced as a dynamic constrain on relative amplitude of signals at IMS stations is characterized by the scattering of station estimates similar to the scattering of network body wave magnitudes. The events built with **RM** are dynamically consistent.

The frequency distribution of **RM** for signals used in all events is exponential. For the IMS stations used in this study, the distributions are similar and also exponential.

## Acknowledgements
The authors are grateful to all analysts at the IDC for reviewing XSEL and REB events. This publication has been produced with the assistance of the European Union, EU Council Decision 2010/CFSP of 26 July 2010.